\documentclass{ws-ijmpcs}

\begin{document}

\markboth{M. Oka}
{Charmed Dibaryon Resonances}

%
\catchline{}{}{}{}{}
%

\title{Charmed dibaryon resonances in the potential quark model}

\author{Makoto Oka}
\address{Advanced Science Research Center, Japan Atomic Energy Agency\\
Tokai, Ibaraki, 319-1195, Japan\\
and\\
Department of Physics, Tokyo Institute of Technology\\
Meguro, Tokyo 152-8551, Japan\\
email: oka@post.j-parc.jp}

\author{Saori Maeda}
\address{Department of Physics, Tokyo Institute of Technology\\
Meguro, Tokyo 152-8551, Japan}

\author{Yan-Rui Liu}
\address{School of Physics, Shandong University\\
Jinan, Shandong 250100, China}

\maketitle

\begin{history}
\published{17 March, 2019}
\end{history}

\begin{abstract}
Charmed dibaryon states with the spin-parity $J^{\pi}=0^{+}, 1^{+}$, and $2^{+}$ are predicted for the two-body
$Y_{c} N$ ($= \Lambda_{c}$, $\Sigma_{c}$, or $\Sigma_{c}^{*}$) systems.
We employ the complex scaling method for the coupled channel Hamiltonian with the
$Y_{c}N$-CTNN potentials, which were proposed in our previous study.
We find  four sharp resonance states near the $\Sigma_{c}N$ and $\Sigma_{c}^{*}N$ thresholds.
From the analysis of the binding energies of partial channel systems, we conclude that these resonance states are Feshbach resonances.
We compare the results with the $Y_{c}N$ resonance states in the heavy quark limit, where the $\Sigma_{c}N$ and $\Sigma_{c}^{*}N$ thresholds are degenerate,
and find that they form two pairs of the heavy-quark doublets in agreement with the heavy quark spin symmetry.
\end{abstract}

\section{Introduction}	
Recent discoveries of new and exotic charmed hadrons have stimulated the interest on interactions of charmed hadrons.
An interesting and not-yet-well-explored subject is the charmed nucleus, {\it i.e.}, a bound state of charmed baryon(s) in nuclei.
As the charm quark is heavier than the strange quark and thus the interaction of charm has novel properties and symmetry,
it is intriguing to explore the interactions and properties of charmed baryons in nuclear medium\cite{DK}.

Many analyses were done for the interaction between the charmed baryon and the nucleon,
from the viewpoint of phenomenological models\cite{Bando}, effective theory\cite{HK}, or lattice QCD\cite{Miyamoto}.
It was pointed out that the attraction between the charmed baryon $\Lambda_c$ and the nucleon is generally weaker than the
corresponding hyperon interactions, mainly because the $K$ meson exchange is replaced by the heavier $D$ meson exchange.
On the other hand, in a recent analysis\cite{ref-liu-oka}, we pointed out that the couplings of the $\Sigma_c$ and $\Sigma_c^*$
baryons, especially through the tensor force, may bring extra attraction.

In a previous study\cite{ref-sm}, we applied a potential model of the $Y_{c}N$ 2-body system,
where $Y_c$ denotes the charmed baryon, $\Lambda_{c}$, $\Sigma_{c}$, or $\Sigma_{c}^{*}$.
We studied the spin-parity $J^\pi=0^+$ and $1^+$ states of $\Lambda_c N$ with full couplings to the corresponding
$\Sigma_c N$ and $\Sigma_c^* N$ channels.
The diagonal and off-diagonal potentials among the channels are calculated in the meson exchange
picture, supplemented by the short-range repulsion taken from the quark exchange model.
The $D$ wave mixings due to the tensor force of the pion exchange potential are taken into account.
We pointed out that a shallow $\Lambda_c N$ bound state may exist both in $J^\pi=0^+$ and $1^+$,
where mixings of $\Sigma_c N$ and $\Sigma_c^* N$ play significant roles.
The near-degeneracy of $J^\pi=0^+$ and $1^+$ states indicates the heavy-quark spin symmetry
in the two-baryon system.

In this study, we extend our analysis at above the $\Lambda_c N$ threshold and study resonance
states around the $\Sigma_c N$ and $\Sigma_c^* N$ thresholds.
It happens  that there exist bound states in the $\Sigma_c N$ (or $\Sigma_c^* N$) single channel
potential, if we omit the coupling to the $\Lambda_c N$ channel.
Then corresponding sharp (Feshbach-type) resonances may appear below the $\Sigma_c N$ (or $\Sigma_c^* N$)
threshold.
We apply the complex scaling method to the coupled channel potential and obtain the resonance energy and width.
(Details are given in Ref.~\refcite{ref-main}.)

\section{Formulation}

\subsection{Channels}
We study the $Y_{c}N$ two-body system with $J^{\pi}=0^{+}$, $1^{+}$, and $2^{+}$ by solving the coupled-channel Schr\"odinger equations.
The list of the coupled channels are given in
Table~\ref{tb:res2-1} for $J^{\pi}=0^+$ and $1^+$ and Table~\ref{tb:res2-2} for $2^+$.
As we are interested in bound or resonance states, we consider the S-wave channels mainly,
but the couplings to the D-wave channels are also taken into account.
The coupling comes from the tensor force due to the pion exchange interaction.
In the S-wave $\Lambda_{c}N$ bound-state calculation, we consider
$0^{+}$ and $1^+$ channels, while
for the $\Sigma_{c}^{*}N$ system, we also consider $2^{+}$ states.

\begin{table}[ht]
\tbl{The S-wave $\Lambda_{c}N$ channels and the channels to couple for $J^{\pi}=0^+$ and $1^+$.}
{\begin{tabular}{ccccccc} \toprule
 $\Lambda_{c}N({}^{1}S_{0})$ & $\Sigma_{c}N({}^{1}S_{0})$ & $\Sigma_{c}^{*}N({}^{5}D_{0})$ &   &   &   &   \\
 \colrule
 $\Lambda_{c}N({}^{3}S_{1})$ & $\Sigma_{c}N({}^{3}S_{1})$ & $\Sigma_{c}^{*}N({}^{3}S_{1})$ & $\Lambda_{c}N({}^{3}D_{1})$ & $\Sigma_{c}N({}^{3}D_{1})$ & $\Sigma_{c}^{*}N({}^{3}D_{1})$ & $\Sigma_{c}^{*}N({}^{5}D_{1})$ \\ \botrule
\end{tabular}
\label{tb:res2-1}}
\bigskip
\tbl{The $J^{\pi}=2^+$ D-wave $\Lambda_{c}N$ and the channels to couple.}
{\begin{tabular}{cccccc} \toprule
 $\Lambda_{c}N({}^{1}D_{2})$ & $\Sigma_{c}N({}^{1}D_{2})$\\ \colrule
 $\Lambda_{c}N({}^{3}D_{2})$ & $\Sigma_{c}N({}^{3}D_{2})$ &
 $\Sigma_{c}^{*}N({}^{3}D_{2})$ & $\Sigma_{c}^{*}N({}^{5}S_{2})$ & $\Sigma_{c}^{*}N({}^{5}D_{2})$ & $\Sigma_{c}^{*}N({}^{5}G_{2})$  \\ \botrule
\end{tabular}
\label{tb:res2-2}}
\end{table}

\subsection{Hamiltonian}
In our previous study\cite{ref-sm}, we introduced a set of two-body coupled channel potentials (called CTNN potential) for the  $Y_{c}N$ system.
The CTNN potentials are composed of the meson exchange potential and the short-range repulsion motivated by the quark exchange dynamics. The former is based on the effective Lagrangian\cite{ref-liu-oka} with the heavy quark symmetry, chiral symmetry, and hidden local symmetry.
The Hamiltonian is given by
\begin{eqnarray}
H^{J} & = & T + V_{\pi} +V_{\sigma} + V_{QCM}, \label{eq:res2-1}\\
V_{\pi}& = & C_{\pi}\frac{m_{\pi}^{3}}{24\pi f_{\pi}^{2}}\left\{ \left< \bm{\mathcal{O}}_{spin} \right> Y_{3}(m_{\pi}, \Lambda_{\pi} ,r) + \left< \bm{\mathcal{O}}_{ten} \right> H_{3}(m_{\pi}, \Lambda_{\pi} ,r) \right\}, \nonumber \\
V_{\sigma}& = & C_{\sigma}\frac{m_{\sigma}}{16\pi}\left\{ 4Y_{1}(m_{\sigma}, \Lambda_{\sigma} ,r) + \left< \bm{\mathcal{O}}_{LS} \right>\left( \frac{m_{\sigma}}{M_{N}} \right)^{2} Z_{3}(m_{\sigma}, \Lambda_{\sigma} ,r) \right\},\nonumber
\end{eqnarray}
where the radial parts, $Y_{1,3}$, $Z_3$ and $H_3$, are the Yukawa potential and its derivatives
convoluted by a form factor designated by the cutoff parameter $\Lambda$.
$C$'s are the coupling constants in each channel with the isospin matrix elements.
All the definitions of the spin operators as well as their matrix elements are given
in Refs.~\refcite{ref-sm} and \refcite{ref-main}.%
\footnote{There is a typo in $V_{\pi}$ of Eq.~(2) in Ref.~\refcite{ref-main}. The $Y_1$ in the first term should be replaced by $Y_3$.}

The last term of Eq.~(\ref{eq:res2-1}) represents the short-range repulsive interaction coming from the quark exchanges between two baryons\cite{ref-2}. We assume the Gaussian form,
\begin{equation}
V_{QCM}=V_{0}e^{-(r^{2}/b^{2})},
\label{eq:res2-2}
\end{equation}
where the strength $V_{0}$ and the range $b$ parameters  are given in Ref.~\refcite{ref-main}.

\subsection{Complex scaling}
To investigate the $Y_{c}N$ resonance states,
we use the Complex Scaling Method\cite{ref-csm1}.
We introduce a complex rotation of the radial coordinate,
$r \rightarrow r e^{i\theta}$ and, correspondingly a rotation of the conjugate momentum, $k \rightarrow k e^{-i\theta}$.
Then we solve the Schr\"odinger equation for the transformed Hamiltonian, 
\begin{eqnarray}
&&{\mathcal H}(r, k) \to  {\mathcal H}(r e^{i\theta}, k e^{-i\theta}) 
\end{eqnarray}
For instance, the kinetic energy term is transformed as
\begin{eqnarray}
&& T = -\frac{1}{2m}(\frac{\partial^{2}}{\partial r^{2}} + \frac{2}{r}\frac{\partial}{\partial r}) \longrightarrow -\frac{1}{2m} e^{-i 2 \theta}(\frac{\partial^{2}}{\partial r^{2}} + \frac{2}{r}\frac{\partial}{\partial r}) \nonumber
\end{eqnarray}

Accordingly the eigen-energy of the Hamiltonian for a scattering, or resonance, state becomes a complex value given by
\begin{eqnarray}
&&  E= e^{-i 2 \theta} \frac{k^2}{2m} 
\nonumber
\end{eqnarray}
where $k$ denotes the asymptotic wave number.
Thus, by choosing appropriate values of $\theta$, we rotate the continuum cut line in the complex plane so that the resonance pole is located above the cut. The corresponding resonance wave function will then decay to zero at large $r$, which
can be obtained in the Gaussian expansion method similarly to the bound state.
Namely, solving the above equation, a resonance state appears at the complex energy
$E = E_R+ i(\Gamma/2)$ as a ``bound'' state,
as far as the condition $2\tan ( 2\theta ) > (\Gamma/E_R)$ is satisfied.
The technical details of the complex scaling method are given in Ref.~\refcite{ref-csm1}.

\section{Results}

\subsection{Resonance energies and widths}
The results are summarized in Table \ref{tb:res2-3}.
In the $J^{\pi}=0^{+}$ channel, there is one resonance state near the $\Sigma_{c}N$ threshold.
The state is very close to the $\Sigma_{c}N$ threshold and has a narrow width.
On the other hand, there is no resonance state around the $\Sigma_{c}^{*}N$ threshold.

In the $J^{\pi}=1^{+}$ channel, there are two resonance states, one each near the $\Sigma_c N$ and $\Sigma_c^* N$ thresholds.
The resonance state near the $\Sigma_{c}N$ threshold has a larger binding energy and a broader width, while
the one near the $\Sigma_{c}^{*}N$ threshold is narrow and has similar properties
as the $0^{+}$ resonance.

In the $J^{\pi}=2^{+}$ channel, there is one resonance near the $\Sigma_{c}^{*}N$ threshold.
The existence of this resonance state is a distinctive characteristic of two-body charmed baryon - nucleon systems,
containing the $\Sigma_{c}^{*}N$ channel because the spin of $\Sigma_c^*$ is $3/2$.
This resonance state has also a large binding energy and a broad width like the resonance state near the $\Sigma_{c}N$ threshold with $J^{\pi}=1^{+}$.

From the study of mixing probabilities of the $\Lambda_c N$, $\Sigma_c N$, and $\Sigma_c^* N$ states obtained (in the bound state approach, where the wave function can be normalized), it is found that the channel coupling is critically important for these resonances.
It is also pointed out that the $D$ wave mixings due to the pion-exchange tensor force are significant for these resonances.
\begin{table}[ht]
\tbl{Energies from the $\Lambda_cN$ threshold ($E_R$), energies from the $\Sigma_cN$ or $\Sigma_c^*N$ threshold ($\Delta E$), and widths ($\Gamma$) of the $Y_{c}N$ resonance states in units of MeV.}
{\begin{tabular}{crcc} \toprule
 states & $E_R$ [MeV] & $\Delta E$ [MeV]&{$\Gamma$ [MeV]} \\
 \colrule
 $\Sigma_{c}N$ $J^{\pi}=0^{+}$ & 163 & $-4$ & 1 \\
 $\Sigma_{c}N$ $J^{\pi}=1^{+}$ & 144 & $-23$ & 12 \\ \colrule
 $\Sigma_{c}^{*}N$ $J^{\pi}=1^{+}$ & 225 & $-7$ & 2 \\
 $\Sigma_{c}^{*}N$ $J^{\pi}=2^{+}$ & 206 & $-25$ & 14 \\ \botrule
\end{tabular}
\label{tb:res2-3}}
\end{table}

In obtaining the above results, we have chosen several complex scaling angles from 0 to 20 degrees and plot all the
eigen-energies of the solutions of the Schr\"odinger equation. These energies are complex but discretized as we
solve the differential equation by expanding the radial wave functions in terms of the Gaussian basis states with
finite extension parameters. Thus the radial wave functions are restricted into a finite volume so that the energies
are discrete. Among the discrete eigenstates, most are scattering states which are aligned along the continuum cut.
The cut line is rotated in the complex energy plane with the change of
the scaling angle $\theta$, and thus the scattering states will move along the threshold line accordingly.

In contrast, the resonance state has a definite pole position in the complex plane so that it stays at the same
point independent of the values of $\theta$.
As an example, we show the eigenvalue plots of the $2^+$ resonance state in Figs.~1 and 2.
One sees that most states are aligned along the lines, which designate the continuum cuts ending at the three energy thresholds, $\Lambda_c N$ (set as $E=0$), $\Sigma_c N$ ($E=167$ MeV) and $\Sigma_c^* N$ ($E=232$ MeV).
 The states within the yellow circle correspond to the Feshbach resonance, which barely moves under the change of $\theta$.
\begin{figure}[ht]
\begin{minipage}{0.5\hsize}
\begin{center}
\includegraphics[width=70mm,bb= 0 0 720 540]{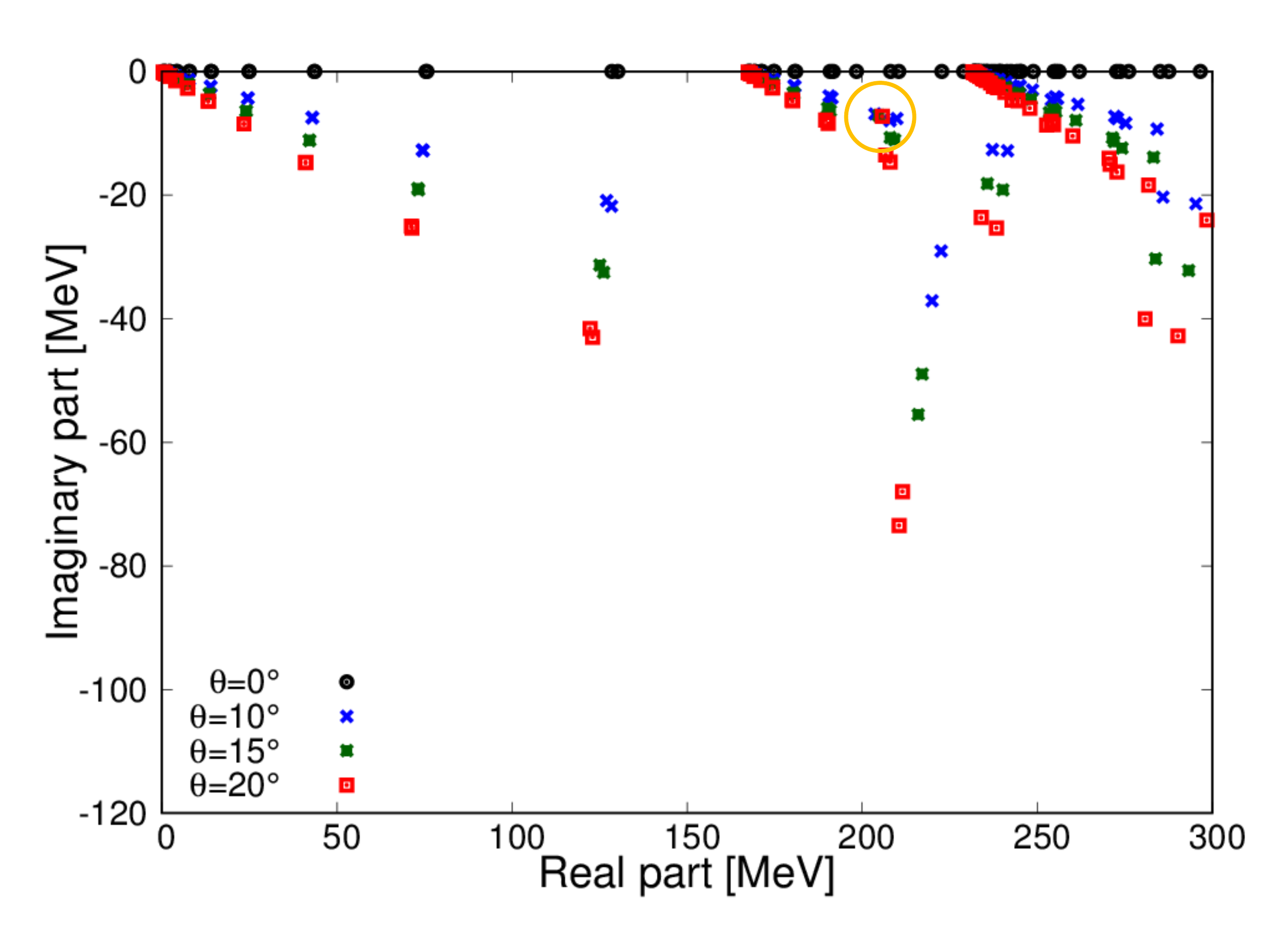}
\caption{\footnotesize{The complex energy eigenvalues for various $\theta$ in the complex scaling method in $J^{\pi}=2^{+}$}.}
\label{gr:res2-6}
\end{center}
\end{minipage}
\hskip 0.5cm
\begin{minipage}{0.4\hsize}
\begin{center}
\includegraphics[width=60mm,bb=0 0 720 540]{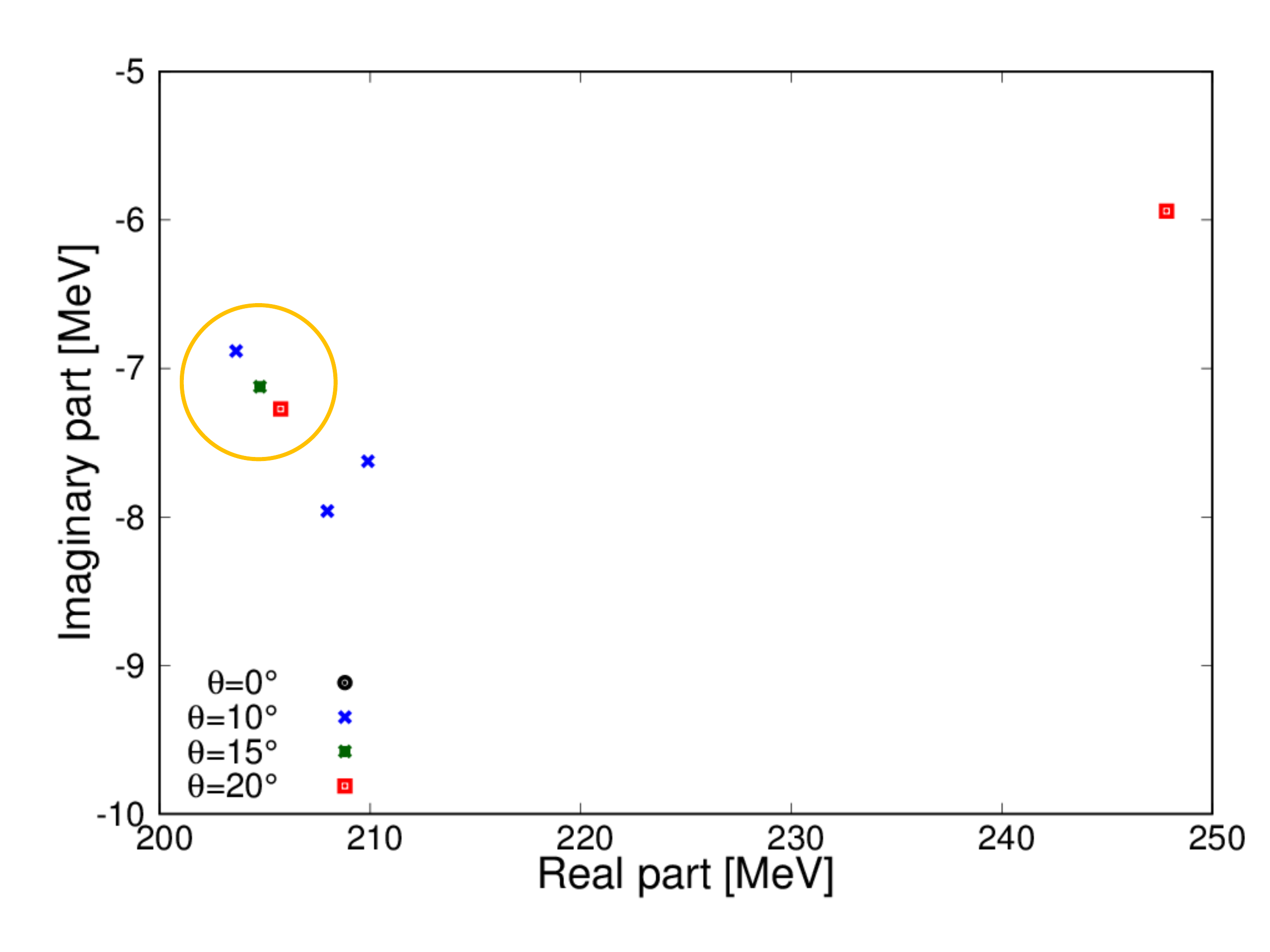}
\caption{\footnotesize{The enlarged view of Fig. \ref{gr:res2-6} focused near the resonance state.}}
\label{gr:res2-7}
\end{center}
\end{minipage}
\end{figure}

\subsection{Heavy quark symmetry}
The obtained resonance states are categorized into two pairs.
One group has narrow widths and another group has broad widths, as shown in Table \ref{tb:res2-3}.
The resonance energy and width of the resonance state near the $\Sigma_{c}^{*}N$ threshold are larger than those of the state near the $\Sigma_{c}N$ threshold.
This pairing property, which is not seen in the strangeness sector, is originated from the heavy-quark spin symmetry.

It is known that the heavy quark spin symmetry of the charm sector predicts heavy quark spin doublets in the single charm hadrons\cite{ref-HQS}.
In particular, $\Sigma_{c}N$ and $\Sigma_{c}^{*}N$ are paired as a heavy quark spin doublet and are degenerate in the heavy quark limit.
To investigate these properties, we calculate resonance states with mass degenerate threshold
in substitution for $\Sigma_{c}N$ and $\Sigma_{c}^{*}N$ thresholds.
The degenerate threshold is obtained from the spin averaged mass, $(m_{\Sigma_{c}N} + 3 m_{\Sigma_{c}^{*}N})/4$.
The heavy mass limit is also taken in the $Y_{c}N$ potentials.
We show the results in Table \ref{tb:res2-6}.
The resonance energies of $\Sigma_{c}N$ increase and those of $\Sigma_{c}^{*}N$ decrease with the degenerate threshold.
One notices immediately that they form two sets of heavy quark doublets.
\begin{table}[ht]
\tbl{Energies from the $\Lambda_cN$ threshold ($E_R$), energies from the $\Sigma_cN/\Sigma_c^*N$ threshold ($\Delta E$), and widths ($\Gamma$) of the $Y_{c}N$ resonance states in units of MeV in the heavy quark limit.}
{\begin{tabular}{ccc|ccc|ccc|ccc} \toprule
 \multicolumn{3}{c|}{$0^{+}$ near $\Sigma_{c}N$}
 & \multicolumn{3}{c|}{$1^{+}$ near $\Sigma_{c}^{*}N$}
 & \multicolumn{3}{c|}{$1^{+}$ near $\Sigma_{c}N$}
 & \multicolumn{3}{c}{$2^{+}$ near $\Sigma_{c}^{*}N$}\\
 $E_R$&$\Delta E$&$\Gamma$&$E_R$&$\Delta E$&$\Gamma$&
 $E_R$&$\Delta E$&$\Gamma$&$E_R$&$\Delta E$&$\Gamma$\\
 \colrule
184 & -5 & 1 & 184 & -5 & 1&
 162 & -27 & 13 & 161 & -28 & 14 \\ \botrule
\end{tabular}}
\label{tb:res2-6}
\end{table}
\begin{figure}[ht]
\includegraphics[width=130mm,bb=0 0 960 425]{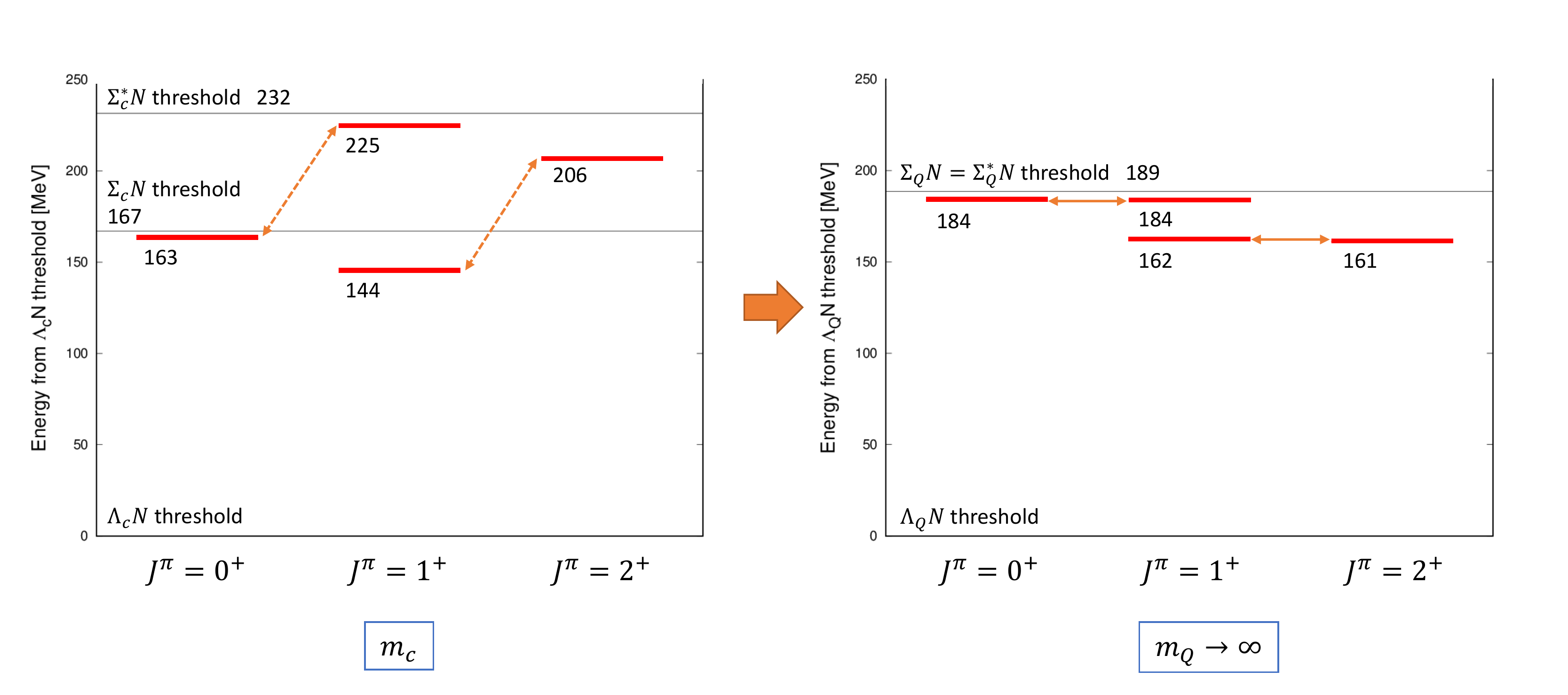}
\caption{Comparison of the resonance energies for the realistic case with for the case in the heavy quark limit.
When $m_{Q} \to \infty$, the thresholds of $\Sigma_{c}N$ and $\Sigma_{c}^{*}N$ will coincide at 189 MeV
above the $\Lambda_c N$ threshold.}
\label{fig:cc-1}
\end{figure}

The comparison of the energy levels for a realistic parameter and the heavy-quark limit are shown in
Figure \ref{fig:cc-1}.
The resonances close to the threshold (5 MeV below) are narrow (width $\sim 1$ MeV), while the deeper ones with
28 MeV below the threshold are broader (width $\sim 14$ MeV)%
\footnote{These values are supposed to be exactly the same for $1^+$ and $2^+$ states, while
the small differences are due to numerical uncertainty.}.
The narrow ones are composed of the charm quark plus light-quark component with the light angular momentum $j=1/2$.
Thus the total angular momentum $J$ is either 0 or 1, which are degenerate in the heavy-quark spin symmetry limit.
Similarly, the broad ones are with $j=3/2$ light-quark component, and thus have total $J= 1$ or 2.
%
%
%

\section{Summary}
\label{sec:res2-6}

We have applied the complex scaling method to the $Y_c N$ potential model in order to explore resonance states
near the $\Sigma_{c}N$ and $\Sigma_{c}^{*}N$ thresholds
in the $\Lambda_c N$, $\Sigma_c N$, and $\Sigma_c^* N$ coupled　channel systems.
Four sharp Feshbach-like resonances are located, one for $J^{\pi} = 0^{+}$, two for $1^{+}$, and one for $2^{+}$.
Each of them is shown to correspond to a bound $\Sigma_c N$ or $\Sigma_c^* N$ state
when we omit the $\Lambda_c N$ channels.
It is found that the $D$ wave mixings due to the pion-exchange tensor force are significant for the resonances.

Comparing the resonance energies from the thresholds, we observe that they form two groups.
The $0^+$ state at the $\Sigma_c N$ threshold and the $1^+$ state at the $\Sigma_c^*N$ threshold
have ``binding'' energies of less than 10 MeV and very narrow widths  ($\Gamma \sim 1-2$ MeV),
while the other two, $1^+$ below $\Sigma_c N$ and
$2^+$ have larger ``binding'' energies ($\sim 25$ MeV) from the threshold and larger widths ($\Gamma\sim 10$ MeV).
These behaviors are consistent with the spin doublet states according to the heavy-quark spin symmetry.

The potential model applied here predicts a shallow bound state of $\Lambda_c N$ in $J^{\pi} = 0^{+}$ and $1^{+}$.
It is extremely interesting to find such bound states in experiment.
However, recent lattice calculation predicts less attractive potential so that no two-body $\Lambda_c N$ bound state may exist.
In the present calculation, among the variations of our potential model, the less attractive one also gives the resonance
states. Thus the searches of the $\Sigma_c N$ and/or $\Sigma_c^* N = \Lambda_c+N+\pi$ resonances should be exciting
even if no $\Lambda_c N$ bound state is found.

\section*{Acknowledgments}

M.O. would like to thank the organizers of the workshop for their hospitality and generous support.
This work is supported in part by JSPS KAKENHI Grant No. 25247036 and by NNSFC Grant No. 11775132.
S. M. was supported by the RIKEN Junior Research Associate Program.


\end{document}